\documentclass[conference]{IEEEtran}
\IEEEoverridecommandlockouts
\usepackage{cite}
\usepackage{amsmath,amssymb,amsfonts}
\usepackage{algorithmic}
\usepackage{graphicx}
\usepackage{textcomp}
\usepackage{xcolor}
\usepackage{multirow}
\usepackage{booktabs}
\usepackage{hyperref}
\usepackage{authblk}
\hypersetup{
    colorlinks=true,
    linkcolor=blue,
    filecolor=magenta,      
    urlcolor=cyan,
    pdfpagemode=FullScreen,
}

\def\BibTeX{{\rm B\kern-.05em{\sc i\kern-.025em b}\kern-.08em
    T\kern-.1667em\lower.7ex\hbox{E}\kern-.125emX}}

\begin{document}

\title{
% FPGA-Based Quad-Camera Visual System for Real-Time Localization in Autonomous Machines 

An Energy-Efficient Quad-Camera Visual System for Autonomous Machines on FPGA Platform 
% quad-camera
% visual frontend
% FPGA
% energy-efficient
% real-time
}

%%%%%%%%%%%%%%%%%%%%%%%%%%%%%%%%%%%%%%%%%%%%%%
% comment out due to double-blind submission %
%%%%%%%%%%%%%%%%%%%%%%%%%%%%%%%%%%%%%%%%%%%%%%

% \author{}
\author[$1$]{Zishen~Wan\textsuperscript{*}}
\author[$2$]{Yuyang~Zhang\textsuperscript{*}\thanks{\textsuperscript{*}Equal Contribution}}
\author[$1$]{Arijit~Raychowdhury}
\author[$3$]{Bo~Yu}
\author[$2$]{Yanjun~Zhang}
\author[$3$]{Shaoshan Liu\thanks{To appear in \textit{IEEE International Conference on Artificial Intelligence Circuits and Systems (AICAS)}, June 6-9, 2021, Virtual}} 

\affil[$1$]{\small School of Electrical and Computer Engineering, Georgia Institue of Technology, Atlanta, GA 30332, USA}
\affil[$2$]{\small School of Information and Electronics, Beijing Institute of Technology, Beijing, 100081, China}
\affil[$3$]{\small PerceptIn Inc, Fremont, CA 94539, USA}

\maketitle

\begin{abstract}
In our past few years' of commercial deployment experiences, we identify localization as a critical task in autonomous machine applications, and a great acceleration target. In this paper, based on the observation that the visual frontend is a major performance and energy consumption bottleneck, we present our design and implementation of an energy-efficient hardware architecture for ORB (Oriented-Fast and Rotated-BRIEF) based localization system on FPGAs. To support our multi-sensor autonomous machine localization system, we present hardware synchronization, frame-multiplexing, and parallelization techniques, which are integrated in our design. Compared to Nvidia TX1 and Intel i7, our FPGA-based implementation achieves 5.6$\times$ and 3.4$\times$ speedup, as well as 3.0$\times$ and 34.6$\times$ power reduction, respectively.
% 67\% and 97\% power reduction, respectively.

\end{abstract}

\section{Introduction}
\label{sec:intro}
% what we did in past few years, visual is important in many scenarios
In the past few years, we have developed and commercialized autonomous machines, such as mobile robots and self-driving cars. During our deployment process, the affordable and reliable \textit{visual frontend} is a critical challenge. With only cameras and IMUs, the visual frontend must precisely perceive the obstacles in unknown environments~\cite{wan2020survey}. An efficient visual system is a prerequisite for localization and exploration tasks.

% sensor fusion for localization, bottleneck, if be accelerated, can benefit many applications
Autonomous machines are complex cyber-physical systems~\cite{krishnan2020sky,krishnan2021machine}. Through detailed performance profiling, the visual frontend is the bottleneck and contributes significantly to system end-to-end latency. Based on our profiling result (Sec.~\ref{subsec:algo_overview}) of the localization task, the vision system accounts for 74\% processing time and consumes more than 50\% CPU resources. Thus, vision frontend is a lucrative acceleration target (Sec.~\ref{subsec:algo_FE}, \ref{subsec:algo_FM}), especially for edge applications with strict real-time and power constraints.

% related work
Several prior works attempt to accelerate visual frontend on low-power platforms. \cite{fang2017fpga} and \cite{cong2011accelerating} implement ORB feature extraction and feature matching on FPGA respectively, but they only accelerate part of the visual system. \cite{suleiman2019navion} designs an optical-flow based VIO (Visual-Inertial Odometry) on ASIC, but optical-flow may fail in variational illuminations and large motion conditions, limiting its application space. \cite{liu2019eslam} presents an ORB-based visual SLAM (Simultaneous Localization and Mapping) design on FPGA, but it only reports results on low-resolution images. Moreover, all of them have not considered large-scale 3D visual systems with multiple cameras, which will provide more robust perception but bring much higher compute intensity and design challenges.

% has varientional delays among different image channels.

% our work
In this paper, we present an FPGA-based quad-camera visual system design for reliable and real-time localization, the system has been commercially deployed in multiple markets around the globe. Specifically, four cameras (with 720p resolution) are integrated into one hardware module, and they create a 360-degree panoramic view of the environment. By utilizing hardware synchronizations (Sec.~\ref{subsec:arch_sync}) and frame-multiplexing processing techniques (Sec.~\ref{subsec:multiplexed}), the most time-consuming feature extraction (Sec.~\ref{subsec:arch_FE}) and feature matching (Sec.~\ref{subsec:arch_FM}) are accelerated on FPGA in a fully pipelined way. Our design achieves 5.63$\times$ and 3.38$\times$ speedup over Nvidia TX1 and Intel i7 CPU, with 3.03$\times$ and 34.63$\times$ power reduction, respectively (Sec.~\ref{sec:eval}). It should be noted that visual odometry should never fail in our design since we can always extract 360-degree spatial information from the environment at any moment, and there are always enough overlapping spatial regions between consecutive frames. Moreover, the efficient use of cameras makes our design much more affordable than LiDAR and its computing systems.

%\begin{figure}[t!]
%        \centering\includegraphics[width=0.75\columnwidth]{figs/camera.png}
%        \vspace{-5pt}
%        \caption{Four-way synchronized images with stereo 360-degree views.}
%        \label{fig:camera}
%        \vspace{-13pt}
%\end{figure}

The main contributions of this paper are as follows:
\begin{itemize}
    \item We propose a novel ORB-based quad-camera 3D visual hardware architecture on an FPGA platform to accelerate the computational-intensive localization frontend.
    
    \item We present a hardware synchronization scheme to support multi-image channels and IMU for reliable localization.
    
    \item We demonstrate how to co-design an accelerator that significantly reduces the latency and energy by exploiting unique frame-multiplexing, parallelisms, and pipelines.
\end{itemize}

\section{Algorithm Framework}
\label{sec:algo}
Hardware design must target critical and compute-intensive blocks. We demonstrate that visual frontend is an universal acceleration target (Sec.~\ref{subsec:algo_overview}), and analyze its two main components, feature extraction (Sec.~\ref{subsec:algo_FE}) and matching (Sec.~\ref{subsec:algo_FM}).

\vspace{-2pt}
\subsection{Visual Frontend Profiling}
\label{subsec:algo_overview}

\begin{figure}[t!]
        \centering\includegraphics[width=.95\columnwidth]{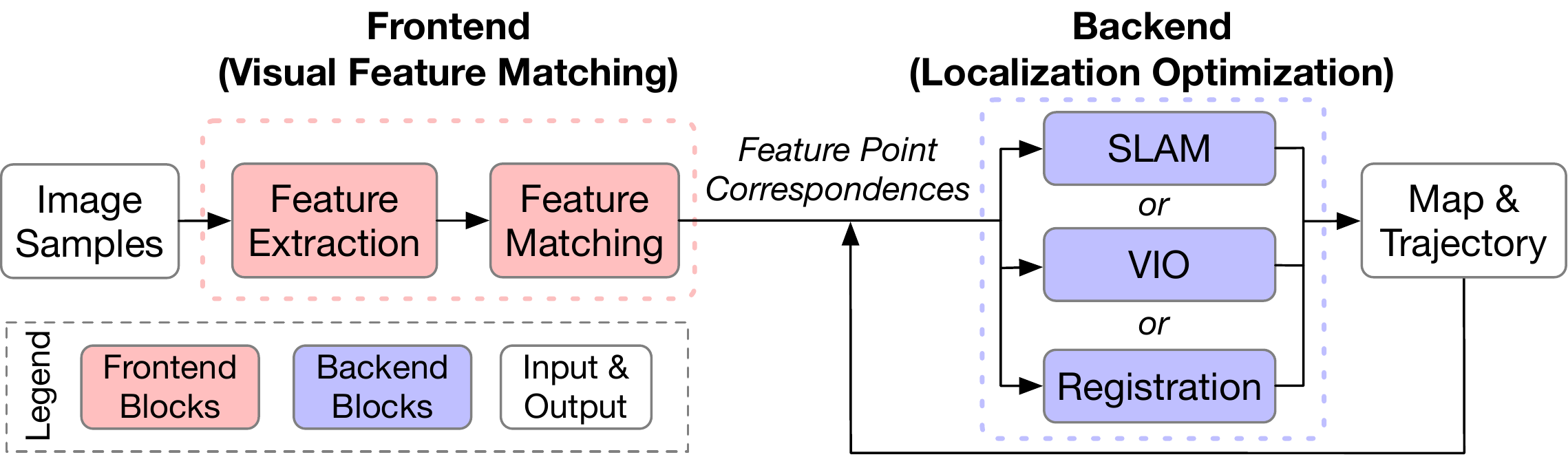}
        \vspace{-5pt}
        \caption{Overview of localization task, including visual frontend and optimization backend.}
        \label{fig:overview}
        \vspace{-13pt}
\end{figure}

To understand the role of visual frontend, we analyze three popular localization algorithms, SLAM~\cite{mur2017orb}, VIO~\cite{forster2016manifold} and Registration~\cite{elbaz20173d}, which are adaptable to different scenarios~\cite{gan2020eudoxus}. As shown in Fig.~\ref{fig:overview}, localization usually consists of a visual frontend and an optimization backend. The frontend extracts visual features to find correspondences in observations, while backend estimates the pose and updates the map. 
All three localization backends share the same ORB-based vision frontend, where ORB has been widely adopted in robotics and it is proven to provide a fast and efficient alternative to SIFT~\cite{lowe1999object}.

Tab.~\ref{tab:profiling} shows the average compute time distribution between the visual frontend and optimization backend of localization systems. We notice that visual frontend is the system bottleneck and its time varies from 54.8\% to 86.7\% in three modes. Therefore, visual frontend is a lucrative acceleration target. Moreover, since different localization algorithms usually use the same image processing approach, accelerating the frontend would lead to a universal performance improvement.

% We measure the compute time of three localization systems in Fig.~\ref{fig:overview}. The average latency distribution between the visual frontend and optimization backend is shown in Tab.~\ref{tab:profiling}. 

The visual frontend consists of feature extraction and feature matching stages (Fig.~\ref{fig:algorithm}). The algorithm details of each stage will be described in Sec.~\ref{subsec:algo_FE} and Sec.~\ref{subsec:algo_FM}, respectively.
% Details will be described as follows.

% The algorithm details of each stage will be described in Sec.~\ref{subsec:algo_FE} and Sec.~\ref{subsec:algo_FM}, respectively.

% We use ORB (Oriented FAST and Rotated BRIEF) to extract features from frames. ORB has been widely adopted in robotics and it is proven to provide a fast and efficient alternative to SIFT~\cite{}. It consists of an FAST (Feature from Accelerated Segment Test)-based feature detector and a BRIEF (Binary Robust Independent Elementary Features)-based feature descriptor. The details are described as follows.

\begin{table}[t!]
\centering
\caption{Latency distribution of frontend and backend in three localization modes.}
\vspace{-5pt}
\resizebox{0.75\columnwidth}{!}{%
\tiny
\begin{tabular}{cccc}
\hline
                  & \textbf{SLAM} & \textbf{VIO} & \textbf{Registration} \\ \hline
\textbf{Frontend} & 54.8\%        & 86.7\%       & 84.6\%                \\
\textbf{Backend}  & 45.2\%        & 13.3\%       & 15.4\%   \\            
\hline   
\end{tabular}
\label{tab:profiling}
}
\vspace{-10pt}
\end{table}

\subsection{Feature Extraction}
\label{subsec:algo_FE}

\begin{figure}[t!]
        \centering\includegraphics[width=\columnwidth]{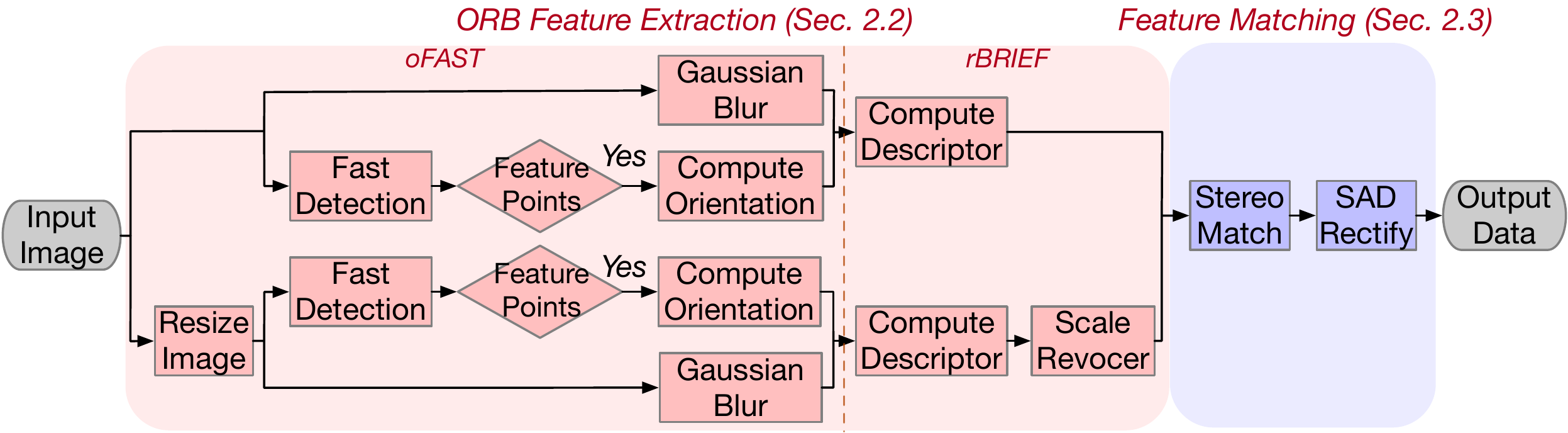}
        \vspace{-16pt}
        \caption{Visual frontend framework, including feature extraction and matching.}
        \label{fig:algorithm}
        \vspace{-13pt}
\end{figure}

We use ORB for feature extraction (Fig.~\ref{fig:algorithm}). ORB is an efficient fusion of FAST (Feature from Accelerated Segment Test) feature detector and BRIEF (Binary Robust Independent Elementary Features) feature descriptor, describing as follows. 
% The details are described as follows.

% \subsubsection{Feature Detection}
% Oriented FAST is used to detect feature points in the image and ensure their rotation invariant. First, the original image is resized to a multi-level image pyramid to enable scale invariance.
% % Second, the feature points are computed at each level of image pyramid, where points  intensity differs greatly to the reference point.
% Second, at each level of image pyramid, points that differ greatly to the reference point in intensity are detected as feature points.
% Third, the orientation of feature points are computed as follows.
\subsubsection{Feature Detection}
Oriented FAST is used to detect feature points in the image and ensure their rotation invariant. First, the original image is resized to a multi-level image pyramid to enable scale invariance.
Second, at each level of the image pyramid, points that differ greatly from the reference point in intensity are detected as feature points.
Third, the orientation of feature points is computed as follows.

Assuming the patch of a feature point is the circle centered at itself, the moments of the patch, $m_{pq}$, are defined as 
\vspace{-6pt}
\begin{equation}
m_{pq} = \sum_{x,y\in{r}}^{} x^p y^q I(x,y), \qquad p,q = 0\;or\;1
\label{eq:mpq}
\vspace{-5pt}
\end{equation}
where $I(x,y)$ is the intensity of the point $(x,y)$ in the patch and $r$ is the patch radius. 
With these moments, the intensity centroid of patch is defined as $C = (m_{10}/m_{00}, m_{01}/m_{00})$,
% \begin{equation}
%     C = (m_{10}/m_{00}, m_{01}/m_{00}),
%     \label{eq:centroid}
% \end{equation}
and the orientation is calculated as $\theta = \text{arctan}(m_{01}/m_{10})$.
% \begin{equation}
%     \theta = \text{arctan}(m_{01}/m_{10})
%     \label{eq:theta}
% \end{equation}

\subsubsection{Feature Description}
Rotated BRIEF computes descriptors of feature points while maintaining their rotation invariance. The detailed steps are as follows. First, considering the circular patch, $p$, a pair of points ($A, B$) is selected, and a binary test, $\tau$, is defined as
\vspace{-6pt}
\begin{equation}
\tau(p;A,B)=
\left\{
             \begin{array}{lr}
             0: p(A) \geq p(B) \\
             1: p(A) < p(B) 
             \end{array}
\right.
\vspace{-5pt}
\end{equation}
where $p(A)$ is the intensity of patch $p$ at point $A$.

Second, $n$ pairs of points are selected from the patch $p$ based on Gaussian distribution. Repeated the previous step $n$ times, the descriptor is calculated as a vector of $n$ binary tests as $f_n(p) = \sum_{1 \leq i \leq n} 2^{i-1} \tau (p; A_i, B_i)$.
% \begin{equation}
%     f_n(p) = \sum_{1 \leq i \leq n} 2^{i-1} \tau (p; A_i, B_i)
% \end{equation}

Third, to ensure rotation invariant, pixels in the patch are rotated by a particular angle around the feature point. In light of the high computation complexity of rotating all points, we choose only to rotate the pairs used for computing descriptors. A $2 \times n$ matrix consisting of these points is defined as
\vspace{-6pt}
\begin{equation}
S = \left(
\begin{matrix}
A_1,A_2,\ldots,A_{n-1},A_{n}\\
B_1,B_2,\ldots,B_{n-1},B_{n}
\end{matrix}
\right)
\vspace{-6pt}
\end{equation}

Using the patch orientation $\theta$ and its rotation matrix $R_{\theta}$, the rotated matrix is derived from $S_{\theta} = R_{\theta}S$. Then the feature descriptor is calculated as $g_n(p,\theta) = f_n(p) | (A_i, B_i) \in S_{\theta}$.
% \begin{equation}
%     g_n(p,\theta) = f_n(p) | (A_i, B_i) \in S_{\theta}
% \end{equation}

\subsection{Feature Matching}
\label{subsec:algo_FM}

\subsubsection{Stereo Matching}
This module matches feature points in a stereo image pair.
First, for a feature point, $F$, in left image, the strip-like searching region $\mathcal{R}$ in right image is determined. Second, the Hamming distance $\mathcal{H}$ of the descriptors between $F$ and each feature point in $\mathcal{R}$ is computed. Third, the feature point pair with the smallest $\mathcal{H}$ is considered as the best match.

\subsubsection{Rectification}
Stereo matching with ORB feature is a local mapping algorithm with fast speed but a high mismatching rate. Thus, SAD (Sum of Absolute Differences) rectification is integrated by correcting the coordinates of feature points. 

Let $(F, F')$ be a matched feature points pair in left and right images. First, patch windows centered at $(F, F')$ are created respectively, as $(F_w, F'_w)$. The SAD is computed as $SAD(F_w, F'_w) = \sum\sum |F_w - F'_w|$, where smaller SAD value means higher similarity of $F$ and $F'$. Second, fix $F_w$ and slide $F'_w$ within a range, repeat the former steps to compute SAD for each new location, find out the window with the lowest SAD, and $F'$ will be relocated to its center. Third, the disparity and depth information is calculated based on adjusted positions.

\section{Hardware Architecture}
\label{sec:arch}
% System overview
% 1. hardware synchronize
% 2. reuse extractor, 2x relation, pipeline
% 3. FE: word length optimization; two-stage shift
% 4. FM: no technique

\begin{figure*}[t!]
        \centering\includegraphics[width=0.88\textwidth]{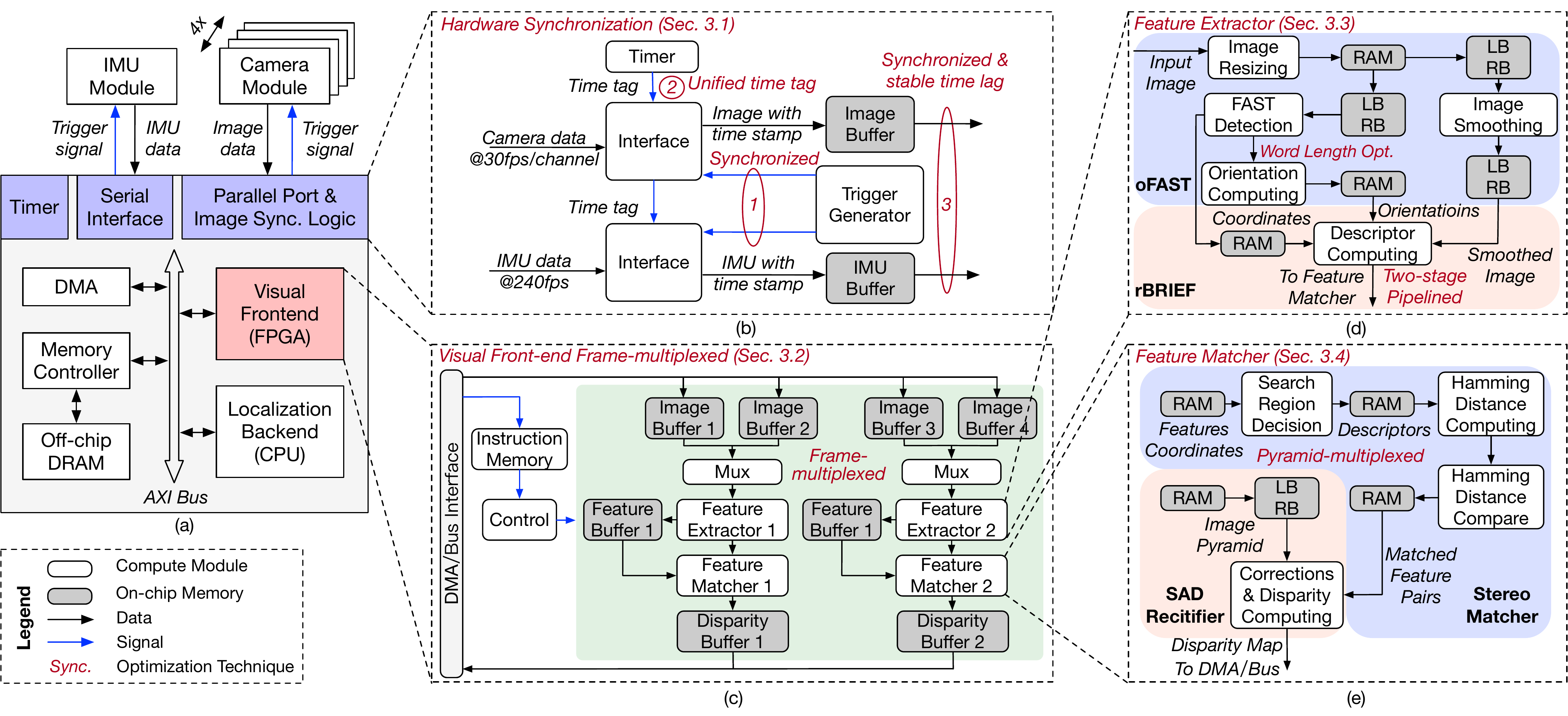}
        \vspace{-0.13in}
        \caption{Overview of proposed FPGA-based hardware architecture design.}
        \label{fig:arch}
        \vspace{-13pt}
\end{figure*}

The overall architecture of the proposed visual frontend accelerator is shown in Fig.~\ref{fig:arch}\textcolor{blue}{a}. Four cameras and IMU interface with on-board compute. The frontend is implemented in FPGA to accelerate feature extraction and matching, and the CPU is used for backend computations. To improve performance, we propose hardware-based synchronization, direct I/O architecture (Sec.~\ref{subsec:arch_sync}) and frame-multiplexed schemes (Sec.~\ref{subsec:multiplexed}). The detailed architecture of feature extraction and matching are presented in Sec.~\ref{subsec:arch_FE} and \ref{subsec:arch_FM}.

% vSLAM Front End Logic:
%  Input: camera images
%  Generic function: calculate features and obtain 3D information  Implementation: FPGA logic
%  Multi-core CPUs:
%  Input: features and 3D information
%  Generic function: multi-view image optimization

\subsection{Hardware Synchronization Interface}
\label{subsec:arch_sync}
% why need hardware synchronization
The synchronized interface between sensors and computing modules is crucial for the autonomous machine to correctly perceive the surroundings and localize itself.
However, in the whole system, the software synchronization in CPU leads to variable delay among the four images, making it impossible to achieve reliable localization results.

% synchronization scheme details
To solve this unstable synchronization issue, we propose a hardware-based synchronization (Fig.~\ref{fig:arch}\textcolor{blue}{b}). First, all captured images are directly sent to on-chip RAM through a direct IO architecture. The trigger generator module generates synchronized trigger signals for cameras and IMU. Second, both input images and IMU data are tagged by a unified time tag. Finally, images and IMU are synchronized at the interface with stable time tags and then sent to the computing modules, significantly helping achieve stable feature processing and localization.
% first, how do you take 4-way camera data and fit it into main computing port, we need to design very special IO (parallel IO) to go into chip. that greatly reduce data movement(bnetweenn IO and camera cause it very slow.)
% aftre that we have synchrinzed iamge go into vSLAM frontend

\vspace{-7pt}
\subsection{Frame-Multiplexed Visual Front-End}
\label{subsec:multiplexed}
% overall structure and profiling
Fig.~\ref{fig:arch}\textcolor{blue}{c} overviews the architecture of vision frontend. 
% The DMA directly accesses data stream captured by cameras via AXI bus and store the data in the image buffers. The results of the feature matchers are finally sent to DDR memory. 
We propose a frame-multiplexed scheme where two camera channels share one feature extractor. The rationale behind this is, based on the profiling result, feature extraction (FE) takes 7.28~\textit{ms} and feature match (FM) takes 14.59~\textit{ms} when processing 640$\times$480 images, indicating that the latency of FM is twice of FE. Moreover, two identical hardware modules are designed to process two stereo cameras in parallel.
% Therefore, we propose a frame-multiplexed scheme where two camera channels share one FE.
%  Two identical hardware modules are designed to process two stereo cameras in parallel.
% to make whole system fully pipelined, which greatly save hardware resources and power. 

\begin{figure}[t!]
\vspace{-5pt}
        \centering\includegraphics[width=0.75\columnwidth]{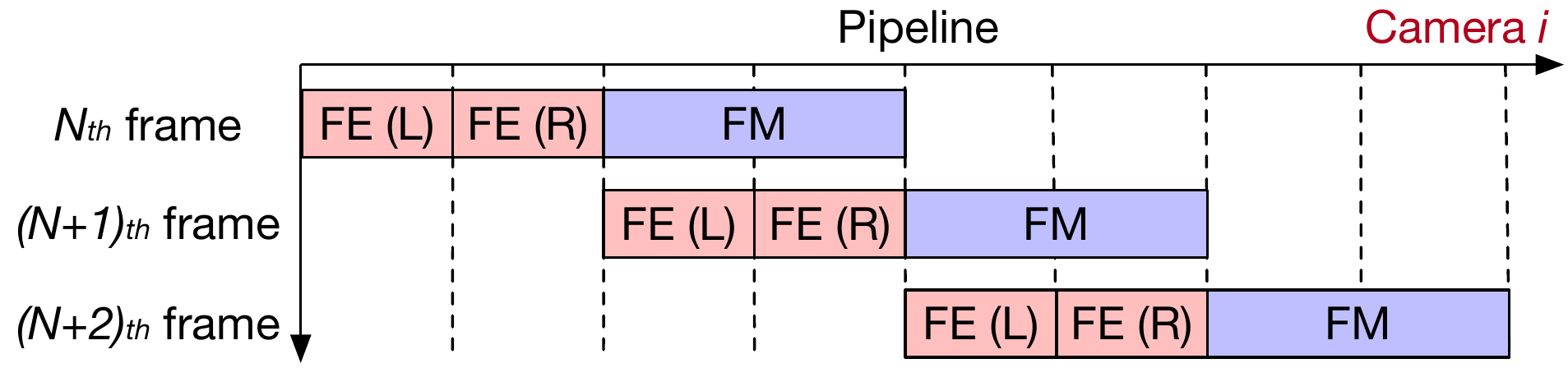}
        \vspace{-5pt}
        \caption{Pipeline in visual front-end, where FE refers to feature extraction, FM refers to feature matching, L/R refers to left/right image. Two camera channels share one FE (frame-multiplexed).}
        \label{fig:pipeline}
        \vspace{-15pt}
\end{figure}

% explain pipeline
The utilized pipeline is shown in Fig.~\ref{fig:pipeline}. Two images (left and right) are captured at each frame. During processing, at $N_{th}$ frame, FE first processes left image and stores the result in buffer and then processes right image. After the extraction is finished, feature descriptors of image pyramid are sent to FM for disparity computation. When FM at $N_{th}$ frame starts to work, FE is fired up to process images for $(N+1)_{th}$ frame.

% pros
With this frame-multiplexed scheme, FE and FM could be performed efficiently in pipelining. Visual frontend runs in parallel with optimization backend. This scheme significantly saves hardware resources and improves throughput.

\vspace{-5pt}
\subsection{Hardware Architecture of Feature Extractor}
\label{subsec:arch_FE}
% input and output
The feature extractor block extracts features from images. It reads data from image buffer and calculates the ORB features with on-chip memory. After the task is finished, it sends the features to buffer and descriptors to feature matcher block. 

% module overview
The detailed architecture of feature extraction is shown in Fig.~\ref{fig:arch}\textcolor{blue}{d}. It consists of image resizing, FAST detection, orientation computing, image smoothing, and descriptor computing modules. RAM, line buffers (LB), and register banks (RB) are used to store intermediate results. The details are as follows.

% intro each module
\textbf{Image Resizing.} 
This module builds a two-layer image pyramid with bilinear interpolation. The size of an input image is 1280$\times$720 and a scaled image is 1067$\times$600. 
% The resized images are stored in RAM.

\textbf{FAST Detection.}
The FAST Detection module takes a 31$\times$31 patch from RB as an input. It detects the feature point and computes the moments of patch (e.g., $m_{10}$ in Eq.~\ref{eq:mpq}). The coordinates of the feature point $(x,y)$ are stored in RAM.
% , and moments are passed to Orientation Computing module.

\textbf{Orientation Computing.} 
This module takes the intensity centroid of a patch as input, and calculates the orientation $\theta$ of the patch that are stored in RAM.

% Since the orientation computation involves division operation, we use 8-bit to represent moments $m_{pq}$ to reduce hardware cost. 

\textbf{Image Smoothing.} 
The image smoothing module utilizes Gaussian filter to smooth 7$\times$7 pixels patch stored in RB. The smoothened images are used to compute feature descriptors.

\textbf{Descriptor Computing.}
This module takes the smoothened images, coordinates and orientations of feature points as inputs, and determines their descriptors with 32$\times$8 bits. 
% To avoid smoothened images occupying too much on-chip memory, we adopt synchronized two-stages shifting line buffers to compute Gaussian filtering and descriptor in a streaming way.

\textbf{Design Techniques.} Orientation computing involves division and square root operations that require substantial costs. We adopt a word length optimization method and choose 8-bit to reduce hardware consumption. During descriptor computation, to avoid smoothened images occupying too much memory, we adopt synchronized two-stages shifting line buffers to compute Gaussian filtering and descriptor in a streaming way.
% design techniques: word length optimization and two-stage pipeline (better to pick it out?)

% rBRIEF algo innovation:
% Since rotating all the points in the patch is computationally expensive, our design only rotates the pairs of points that are used for computing descriptors.
\vspace{-4pt}
\subsection{Hardware Architecture of Feature Matcher}
\label{subsec:arch_FM}
% input and output
The feature matcher block aims to match feature points from stereo images and derive depth information. It contains two parts in our design, the stereo matcher for pre-match and the SAD rectifier for further rectification.

% module overview
The detailed architecture of feature matcher is demonstrated in Fig.~\ref{fig:arch}\textcolor{blue}{e}, including region decision, distance computing and compare, correction and disparity computing modules. The design of each module is presented as follows.

% intro each module
\textbf{Search Region Decision.}
Search Region Decision module takes the coordinates of feature points as inputs and determines whether the feature points locate within the searching field. 

\textbf{Distance Computing and Compare.}
This module obtains the descriptors from RAM and computes the Hamming distance of each pair of feature descriptors. It then finds out the best matching with the smallest value of Hamming distance.
% \textbf{Distance Compare.}

\textbf{Correction and Disparity Computing.}
This module is used for SAD rectification. It takes the coordinates of matched feature pairs obtained in stereo matcher and 11$\times$11 patch image pyramid from RB, and computes the corrections and disparity. The depth information will be sent for backend use.

% sent to AXI bus for backend use.
\textbf{Design Techniques.} We utilize an image pyramid-multiplexed scheme during implementation where two resizing images share the same Feature Matcher block, significantly saving hardware resources costs.
\section{Evaluation Results}
\label{sec:eval}

% 1. setup
% 2. resource
% 3. accuracy (path)
% 4. power and performance
\vspace{-2pt}
\subsection{Experimental Setup}
\label{subsec:exp_setup}
\textbf{Hardware Platform.}
The proposed visual front-end accelerator is implemented and evaluated on Xilinx Zynq Ultrascale+ XCZU9EG MPSoC. The FPGA is directly interfaced with the four cameras and IMU sensors. The max operated clock frequency is 203 MHz for feature extractor and 230MHz for feature matcher, respectively. The FPGA device has 274K LUTs, 548K Flip-Flops, 912 BRAMs, and 2520 DSPs in total.

\begin{table}[t!]
\caption{FPGA resource consumption of system.}
\vspace{-5pt}
\resizebox{\columnwidth}{!}{%
\begin{tabular}{c|ccc|cc}
\toprule[0.6pt]
\multirow{2}{*}{\textbf{Resource}} & \multicolumn{3}{c|}{\textbf{Modular Used (640$\times$480)}} & \multicolumn{2}{c}{\textbf{Total Used}} \\ \cline{2-6} 
                          & \textbf{FE}      & \textbf{FM}      & \textbf{Ctrl.}      & \textbf{640$\times$480 }      & \textbf{720$\times$1280}       \\ \hline
LUT                       &    96850      &   40034      &   1759           &      138643 (51\%)         &      177196 (65\%)          \\
Flip-Flop                 &    54100     &  12694       &    479          &      67273 (12\%)         &       82730 (15\%)         \\
BRAM                      &    271      &   0      &     0         &    271 (30\%)           &    785 (86\%)             \\
DSP                       &    32      &   0      &      0        &    32 (1\%)           &   109 (4\%)  \\          
\bottomrule[0.6pt]
\end{tabular}
\label{tab:resource}
}
\vspace{-10pt}
\end{table}

\textbf{Resource Consumption.} 
The resource consumption of the proposed system is shown in Tab.~\ref{tab:resource}. We evaluate the design on two different image resolutions. 
% The whole 4-channel 3D-vision frontend system is composed of 2 FEs with frame multiplexing (Sec.~\ref{subsec:multiplexed}), 2 FMs and external control logic with additional memory. 
Overall, the hardware architecture utilizes 51\% LUT, 12\% Flip-Flop, 30\% BRAM, and 1\% DSP resources when processing 640$\times$480 images.

Particularly, in 640$\times$480 images, it is observed that FE consumes over two-thirds of the frontend resource; the percentages in 720$\times$1280 images are similar, corroborating our scheme to multiplex FE module between left and right frames.

% It is well observed that frontend dominates the resource consumption. The FE and FM modules consume xx\% LUT, xx\% Flip-Flop, xx\% DSP and xx\% BRAM of the total used resources. Particularly, FE consumes two-thirds of the frontend resource, corroborating our scheme to multiplex FE module between left and right frames.

% The whole 4-channel 3D-vision front-end system is composed of 2 feature extractors with frame multiplexing (Sec.~\ref{subsec:multiplexed}), 2 feature matchers and external control logic with additional memory. Overall, the hardware architecture utilizes xx\% LUT, xx\% FF, xx\% BRAM and xx\% DSP resources.

\vspace{-4pt}
\subsection{Accuracy Analysis}
\label{subsec:accuracy}

\begin{table}[]
\caption{Accuracy Evaluation of FPGA System.}
\vspace{-5pt}
\resizebox{\columnwidth}{!}{%
\begin{tabular}{cccc}
\toprule[0.6pt]
                  & \textbf{\# Feature Points} & \textbf{\# Matched Pairs} & \textbf{\# Effective Depth Value} \\ \hline
\textbf{Software} & 961.2                           & 211.5                                         & 107                                        \\
\textbf{FPGA}     & 961.1                           & 211.7                                         & 107.3                                      \\
\textbf{Error}    & -0.1                            & +0.2                                          & +0.3                  \\              \bottomrule[0.6pt]      
\end{tabular}
\label{tab:accuracy}
}
\vspace{-13pt}
\end{table}

% \begin{figure}[t!]
%         \centering\includegraphics[width=0.75\columnwidth]{figs/match_result.pdf}
%         \caption{xx.}
%         \label{fig:result_match}
% \end{figure}

The accuracy of the system is evaluated by the results of feature extraction and matching between FPGA and software implementation (MATLAB) with processing 30 frames, shown in Tab.~\ref{tab:accuracy}.
% the number of extracted feature points, matched feature points pairs and obtained depth values of two approaches are almost the same (error$<$0.3\%). 
For the number of extracted feature points, matched feature points pairs and obtained depth values, the results of two approaches are almost the same (error$<$0.3\%).
For detailed coordinates, the accuracy is 99.7\%, 98.2\% and 96.8\%. 

\vspace{-4pt}
\subsection{Performance and Power Evaluation}
\label{subsec:performance}

\begin{table}[]
\caption{Performance and Power Comparison.}
\vspace{-5pt}
\resizebox{\columnwidth}{!}{%
\begin{tabular}{ccccc}
\toprule[0.6pt]
% \hline
                       & \textbf{\begin{tabular}[c]{@{}c@{}}Perform. (\textit{fps})\end{tabular}} & \textbf{\begin{tabular}[c]{@{}c@{}}Perform. (\%)\end{tabular}} & \textbf{\begin{tabular}[c]{@{}c@{}}Power (\textit{W})\end{tabular}} & \textbf{\begin{tabular}[c]{@{}c@{}}Power (\%)\end{tabular}} \\ \hline
\multicolumn{5}{c}{Image Resolution: 640x480}                                                                                                                                                                                                                                                      \\ \hline
\textbf{Our}           & 69                                                                   & -                                                                   & 1.63                                                         & -                                                             \\
\textbf{\cite{liu2019eslam}}       & 56                                                                   & 81.16\%                                                             & 1.94                                                         & 1.19$\times$                                                         \\
\textbf{\cite{fang2017fpga}}     & 67                                                                   & 97.1\%                                                              & 4.56                                                         & 2.80$\times$                                                         \\ \hline
\multicolumn{5}{c}{Image Resolution: 1280x720}                                                                                                                                                                                                                                                     \\ \hline
\textbf{Our}           & 50.7                                                                 & -                                                                   & 2.31                                                         & -                                                             \\
\textbf{Nvidia TX1}    & 9                                                                    & 17.75\%                                                             & 7                                                            & 3.03$\times$                                                         \\
\textbf{Intel i7 Core} & 15                                                                   & 29.59\%                                                             & 80                                                           & 34.63$\times$    \\                                                  
\bottomrule[0.6pt]
\end{tabular}
\label{tab:performance}
}
\vspace{-13pt}
\end{table}

\textbf{CPU/GPU Comparison.} Tab.~\ref{tab:performance} compares the performance and power of our FPGA design, Nvidia TX1 and Intel i7 CPU. Compared with TX1, the performance is raised by 5.63$\times$ and power is reduced by 67\%. Compared with i7 CPU, the performance is raised by 3.38$\times$ and power is reduced by 97\%.
% our design achieves 5.63$\times$ speedup and 3.03$\times$ energy efficiency improvement. Compared with i7 CPU, our design achieves 3.38$\times$ speedup and 34.63$\times$ energy efficiency improvement. 

\textbf{Existing Accelerator Comparison.}
% To our best knowledge, no work related to hardware 3D-vision algorithm was published before, so a fair comparison is difficult. 
Tab.~\ref{tab:performance} compares our proposed hardware with some existing accelerators. Compared to~\cite{fang2017fpga} (FPGA), we achieve a slight 1.03$\times$ speedup but 64\% power reduction. Particularly, we propose extra stereo match modules for depth information, which is vital in 3D environments. Compared to~\cite{liu2019eslam} (FPGA), our design achieves 1.23$\times$ speedup and saves 16\% power. Moreover, we support multi-channel camera systems with proposed hardware synchronization scheme. Compared to~\cite{suleiman2019navion} (ASIC), our design achieves 69 \textit{fps} on more robust feature-based stereo-flow method, whereas they achieves 171 \textit{fps} on optical-flow method that may fail in large displacements or inconsistent illumination conditions.

\vspace{-4pt}
\subsection{Discussions}
\label{subsec:discussion}
Our proposed design can be further improved in two directions. For high-performance scenarios, it can be improved by involving more aggressive pipelining and higher fan-out nets reduction. For embedded scenarios, power can be further reduced by optimizing supply voltage and clock frequency.

% Frontend dominates the resource consumption. In EDX-CAR, the frontend uses 83.2\% LUT, 62.2\% Flip- Flop, 80.2\% DSP, and 73.5\% BRAM of the total used resource; the percentages in EDX-DRONE are similar. In particular, feature extraction consumes over two-thirds of frontend resource, corroborating our design decision to multiplex the feature extraction hardware between left and right camera streams

% FF for computing, BRAM for storage, DSP for acceleration 
\vspace{-3pt}
\section{Conclusion}
\label{sec:conclusion}
In this paper, the unified compute bottleneck of various localization system is identified. An ORB-based visual frontend architecture is presented for real-time and energy-efficient localization and evaluated on FPGA platform. To support multiple cameras and IMU, we propose hardware synchronization for stable localization. To accelerate feature extraction and matching, we utilize frame-multiplexing, parallelism and word length optimization. Compared with Nvidia TX1 and Intel i7, our design achieves 5.6$\times$ and 3.4$\times$ speedup in frame rate, and 3$\times$ and 34.6$\times$ improvement in energy efficiency, respectively.
\vspace{-15pt}
\section*{Acknowledgements}
This work was supported in part by C-BRIC, one of six centers in JUMP, a Semiconductor Research Corporation (SRC) program sponsored by DARPA.

\bibliographystyle{ieeetr}
\bibliography{refs}
%%%%%%%%%%%%%%%%%%%%%%%%%%%%%%%%%%%%

\end{document}